\documentclass[12pt,preprint]{aastex}
\usepackage{epsfig}

\begin{document}

\title{Gamma-Ray Constraints on Neutralino Dark Matter Clumps in the
Galactic Halo}

\author{Roberto Aloisio}
\affil{INFN - Laboratori Nazionali del Gran Sasso\\
SS. 17bis, Assergi (AQ), ITALY\\
and\\
Center for Cosmological Physics \\
The University of Chicago, Chicago, IL 60637, USA}
\email{roberto.aloisio@lngs.infn.it}
\author{Pasquale Blasi}
\affil{INAF - Osservatorio Astrofisico di Arcetri\\
Largo E. Fermi, 5 - 50125 Firenze, ITALY\\
and\\
Center for Cosmological Physics \\
The University of Chicago, Chicago, IL 60637, USA}
\email{blasi@arcetri.astro.it}
\author{Angela V. Olinto}
\affil{Department of Astronomy \& Astrophysics, \& Enrico Fermi
Institute, \\
The University of Chicago, Chicago, IL 60637, USA\\
and\\
Center for Cosmological Physics \\
The University of Chicago, Chicago, IL 60637, USA}
\email{olinto@oddjob.uchicago.edu}

\begin{abstract}
According to high resolution cold dark matter (CDM) simulations, large
virialized halos are formed through the constant merging of smaller halos
formed at earlier times. In particular, the halo of our Galaxy may
have hundreds of dark matter clumps. The annihilation
of dark matter particles such as the neutralino in these clumps generates
$\gamma$-ray fluxes that can potentially be detected by future experiments
such as GLAST. We find that, depending on the parameters of the clump density
profile and on the distribution of clumps in the Galactic halo, the
contribution to the diffuse $\gamma$-ray background from clumps can
constrain the properties of neutralinos such as the mass and annihilation
cross section.  We model the density profile of clumps by three
representative  dark matter profiles: singular isothermal spheres (SIS),
Moore profiles, and Navarro, Frenk and White (NFW) density profiles and
calculate  the spectrum and angular distribution in the sky of the
$\gamma$-ray  flux due to neutralino annihilation in the clumpy halo of the
Galaxy.  The calculations are carried out in the context of two different
scenarios for the distribution of clumps in the Galaxy and their
concentrations, which  result in very different conclusions.
\end{abstract}


\section{Introduction}

Most of the matter in the universe has yet to be observed in any
frequency band, thus the name, dark matter (DM). The evidence for the
predominance of  dark over visible matter comes mainly from the
gravitational effects of the dark  matter component. However,
gravitational studies have been unable to shed light on the nature of
the dark matter. Big bang nucleosynthesis constrains  most of the dark
matter to be of non-baryonic origin. This has encouraged the study of
plausible  new particle candidates for the dark matter. The best motivated
among these candidates is the lightest supersymmetric particle, named the
neutralino, $\chi$. Given the requirements of neutralino production in the
early universe, it is possible to study the phenomenology of such dark
matter candidates in detail (see, e.g., Jungman, Kamionkowsky, and Griest
1996).  In particular, the annihilation of  neutralinos has often been
considered  a potential source of detectable secondaries: high energy
particles and  electromagnetic radiation. In this sense, dark matter can be
visible  through the radiation caused by the annihilation secondaries
(Berezinsky, Gurevich and Zybin 1992; Berezinsky, Bottino and Mignola 1994;
Gondolo and Silk 1999; Bergstr\"om et al. 1999; Gondolo 2000;
Calc\'{a}neo--Rold\'{a}n and Moore 2000; Bertone, Sigl and Silk 2001;
Bergstr\"om, Edsj\"o and Ullio 2001; Blasi, Olinto and Tyler 2002; Tyler 2002;
Ullio et al. 2002; Tasitsiomi and Olinto 2002; Cesarini et al. 2003).

Recent advances in cold dark matter (CDM) simulations have shown that the
large scale structure of the universe can be explained in terms of a
hierarchical scenario in which large halos of dark matter are
generated by the continuous merging of smaller halos
(Ghigna et al. 1998; Moore et al. 1999; Klypin et al. 1999).
In this picture, a dark matter halo is the
superposition of a smooth component on a scale comparable with the virial
radius of the forming  structure and a clumped structure made
of thousands of smaller mass halos. As we discuss below, this scenario
can have important consequences for the detection of signals from
neutralino-antineutralino ($\chi\bar{\chi}$) annihilation.

CDM simulations also show that most halos are well described by a
density distribution with cusps at the center of each
halo.  The exact shape of the central cusp is still a matter of debate.
Most recent simulations favor profiles with density cusps varying from
the  Moore et al. profile (Moore et al. 1999) where $\rho_{DM} (r \to 0)\sim
r^{-1.5}$ to a Navarro, Frenk, and White (NFW) profile
(Navarro, Frenk and White 1996) where $\rho_{DM} (r \to 0)
\sim r^{-1}$ (see also Power et al. 2002). The more concentrated the central
cusp, the easier to detect annihilation products since the annihilation rate
is proportional to the square of the dark matter density. The existence
of cuspy halos is still unclear since observations of
galaxy rotation curves provide no evidence  of central cusps
(see, e.g., Salucci 2001). However, the survival of  cusps in galactic
centers is highly dependent on the galaxy's merger  history in particular on
the history of formation of galactic center  black holes (Merrit and
Milosavljevic 2002).  The central regions of small mass dark matter halos
(i.e., DM clumps) are less affected by the dynamics of baryonic matter and
less likely to have black hole mergers at their centers. Thus, a cuspy
profile may well describe the density of DM clumps.

Another important piece of information is embedded in the spatial distribution
and survival history of clumps on their way to the central part of the
host galactic halo. Tidal interactions significantly affect clump
distribution and structure and these effects are only beginning to be
resolved in CDM simulations for relatively large mass clumps (Stoehr
et al. 2002).
Given the lack of a clear understanding of present distribution of
DM clumps in the Galaxy halo, we consider two  different scenarios that
bracket the present uncertainties. For simplicity we call these two
approaches type I and type II. We show that the present uncertainties lead to
large variations in the resulting gamma-ray signal.

In the type I scenario, the clump position in the host Galaxy determines its
external radius. In the case of NFW and Moore profiles, the location of the
core is then determined by assuming a fixed fraction of this external radius.
In the type II scenario, the concentration parameter, defined as the ratio of
the core radius to virial radius of a clump, is modeled as in Wechsler et
al. (2001) for clumps for a given mass. We show that in this scenario the
clumps are  gradually destroyed on their way to the center of the Galaxy,
such that the inner part of the Galactic halo is depleted of clumps. This
second scenario seems compatible  with numerical simulations (Power et al.
2002, Stoehr et al. 2002). The uncertainties on clump distribution and
concentration bracketed by these two scenarios strongly affect the resulting
$\gamma$-ray emission from the annihilation of neutralinos: while in the
type I scenario, the highly concentrated  clumps produce strong
$\gamma$-ray emission, in the  type II scenario the low concentration
clumps imply $\gamma$-ray fluxes several orders of magnitude smaller.

The paper is structured as follows: in section 2 we introduce
the clump density profiles; in section 3 we discuss the distribution of
clumps in the Galaxy and the calculations of the gamma ray flux;
in section 4 we present our results, consisting of the spectra and
anisotropies of the $\gamma$-ray emission in the type I and II scenarios,
for different shapes of the clumps density profiles. We conclude in
section 5.

\section{Dark Matter Clumps in the Halo}

The contribution to the diffuse $\gamma$-ray background
from $\chi\bar{\chi}$ annihilation in the clumpy halo depends on the
distribution of clumps in the halo and on the density profile of clumps.
Our purpose here is to investigate the wide variety of possibilities
currently allowed by the results of simulations and suggested by some
theoretical arguments, concerning the density profiles of dark matter
clumps. We consider three cases: singular isothermal spheres (SIS),
Moore et al. profiles, and NFW profiles.

The Moore et al. and the NFW profiles are both the result of fits
to different high resolution simulations (Moore et al. 1999,
Power et al. 2002). Although there is ongoing debate over which profile
is most accurate, it is presently believed that a realistic descriptions
of the dark matter distribution in halos will follow a profile in the
range defined by the Moore et al. and the NFW fits (see, e.g.,
Tasitsiomi 2002).

The dark matter density profiles can be
written as follows:
\begin{equation}
\rho_{\chi,{\rm SIS}}(r) = \rho_0 \left({r \over r_0}\right)^{-2} .
\label{SIS_profile}
\end{equation}
\begin{equation}
\rho_{\chi,{\rm Moore}}(r) = \frac{\rho_0}
{\left(r \over r_f \right)^{3/2} \left [1+
\left(r \over r_f \right)^{3/2} \right]}
\label{Moore_profile}
\end{equation}
\begin{equation}
\rho_{\chi,{\rm NFW}}(r) = \frac{\rho_0}
{\left(r \over r_f \right) \left(1+
\frac{r}{r_f} \right)^2}~.
\label{NFW_profile}
\end{equation}
The SIS and Moore et al. clump density profiles are in the form given in
eqs. (\ref{SIS_profile}) and (\ref{Moore_profile}) down to a minimum radius,
$r_{\rm min}$.
Inside $r_{\rm min}$, neutralino annihilations are faster than the cusp
formation rate, so that $\rho (r \le r_{\rm min}) = \rho (r_{\rm min})$
remains constant. To estimate $r_{\rm min}$, following Berezinsky, Gurevich
and Zybin 1992, we set the annihilation timescale equal to the free-fall
timescale, so that
\begin{equation}
r_{\rm min,SIS} = r_{0} \left(
\frac{\langle \sigma v \rangle \rho_0}
{\sqrt{G M_c} m_{\chi}} r_0^{3/2} \right)^{1/2} ,
\label{Rmin_SIS}
\end{equation}
\begin{equation}
r_{\rm min,Moore} = r_{0} \left(
\frac{\langle \sigma v \rangle^2 \rho_0^2}
{G M_c m_{\chi}^2} r_f^{3} \right)^{1/3} ,
\label{Rmin_Moore}
\end{equation}
where $\langle \sigma v \rangle_{\chi \bar{\chi}}$ is the ${\chi
\bar{\chi}}$ annihilation cross-section, $G$ is Newton's constant, and
$M_c$ and $m_{\chi}$ are respectively the clump mass and the neutralino
mass.

We modeled the smooth galactic halo density with a NFW
density profile (eq. (\ref{NFW_profile})), with $r_f=27$ kpc and
$\rho_0$ determined by the condition that the dark matter density at
the Sun's position be $\rho_{DM}(d_{\odot}) = 6.5 \times 10^{-25} {\rm
g/cm}^3$.

The fundamental parameters of the clump density profile are the density
normalization $\rho_0$, the clump radius $r_0$ and, in the case of Moore
et al. and NFW profiles, the clump fiducial radius $r_f$. In order to fix
these fundamental parameters we have considered two different scenarios
(type I and II).

In the type I scenario, the radius of a clump with fixed mass is determined
by its position in the Galactic halo. More specifically, the radius of the
clump is located at the radius where the clump density equals the density
of the Galactic (smooth) dark matter halo at the clump position (namely
$\rho_0$). The physical motivation for such a choice is to account for the
tidal stripping of the external layers of the clump while the clump is
moving in the potential of the host halo. For the NFW and Moore profiles,
the fiducial radius $r_f$ has been taken as a fixed fraction of $r_0$:
$r_f=0.1 r_0$.

\begin{figure}[t!]
\plotone{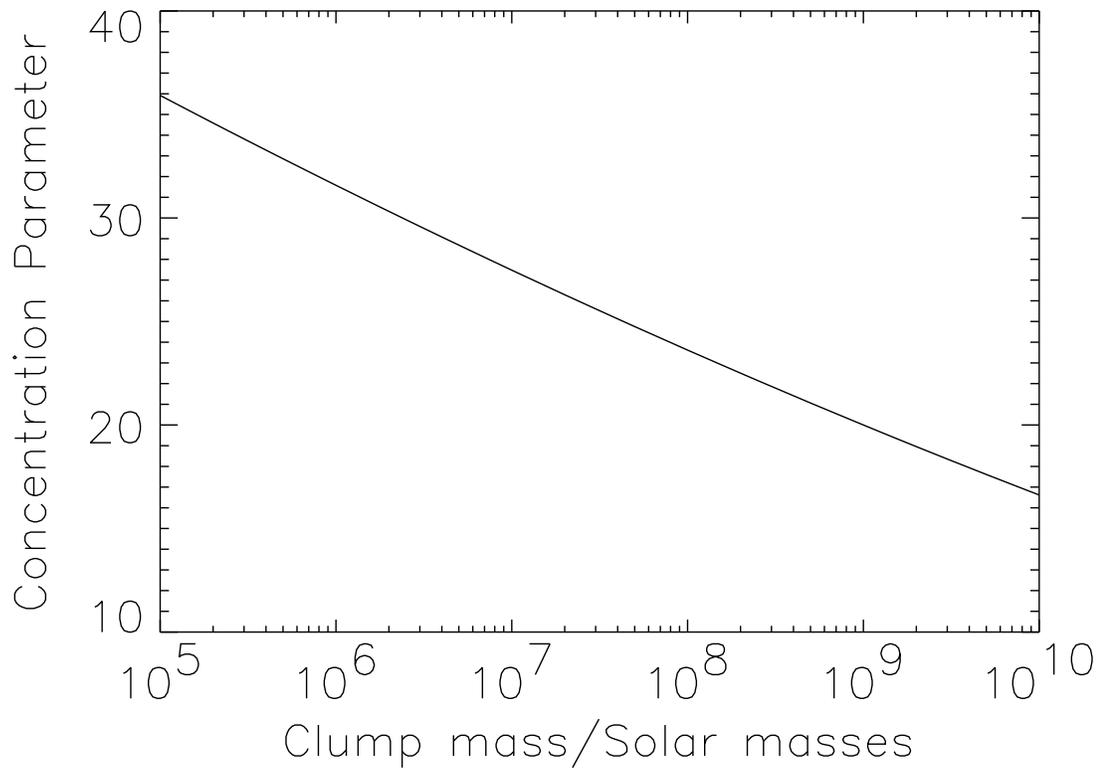}
\caption{Concentration parameter for dark matter clumps, as a function of
the clump mass.}
\label{fig:conc}
\end{figure}

In the type II scenario the external radius of the clumps is taken to
be their virial radius, defined in the usual way:
\begin{equation}
r_0=r_{vir}=\left(\frac{3 M_c}{4\pi\rho_{200}} \right)^{1/3} \ ,
\label{virial}
\end{equation}
where $\rho_{200}$ is 200 times the critical density of the Universe
$\rho_c=1.88\times 10^{-29}h^2$ g/cm$^3$ (we assume $h=0.7$).
In this scenario, following Ullio et al. 2002, we have introduced
the concentration parameter defined as
\begin{equation}
\xi=\frac{r_0}{r_{-2}}
\label{concentr_2}
\end{equation}
where $r_{-2}$ is the radius at which the effective logarithmic slope of
the profile is $-2$, set by the equation
$$ \frac{1}{\rho_0}\frac{d}{d r}r^2\rho(r) = 0~. $$
The mass dependence of the concentration parameter used in our calculations
is taken from Wechsler et al. 2001, and is plotted in fig. \ref{fig:conc} for
the present time (zero redshift). The general trend is for smaller clumps to
have larger concentration parameter, reflecting the fact that they were formed
at earlier epochs, when the universe was denser. 

\clearpage

\clearpage

The normalization constant in the clump density profile $\rho_0$ is fixed
recalling that the clump mass $M_c$ is
$$ M_c = \int_{0}^{r_0} 4\pi r^2 \rho_{\chi}(r) dr~. $$
In the case of the NFW density profile, $r_f=r_{-2}$, while in the case of the
Moore et al. density profile, $r_f=r_{-2}/0.63$ (Ullio et al. 2002). In
terms of concentration parameters
\begin{equation}
\xi_{NFW}=0.63 \, \xi_{Moore}~\ .
\end{equation}

\begin{figure}[t!]
\plotone{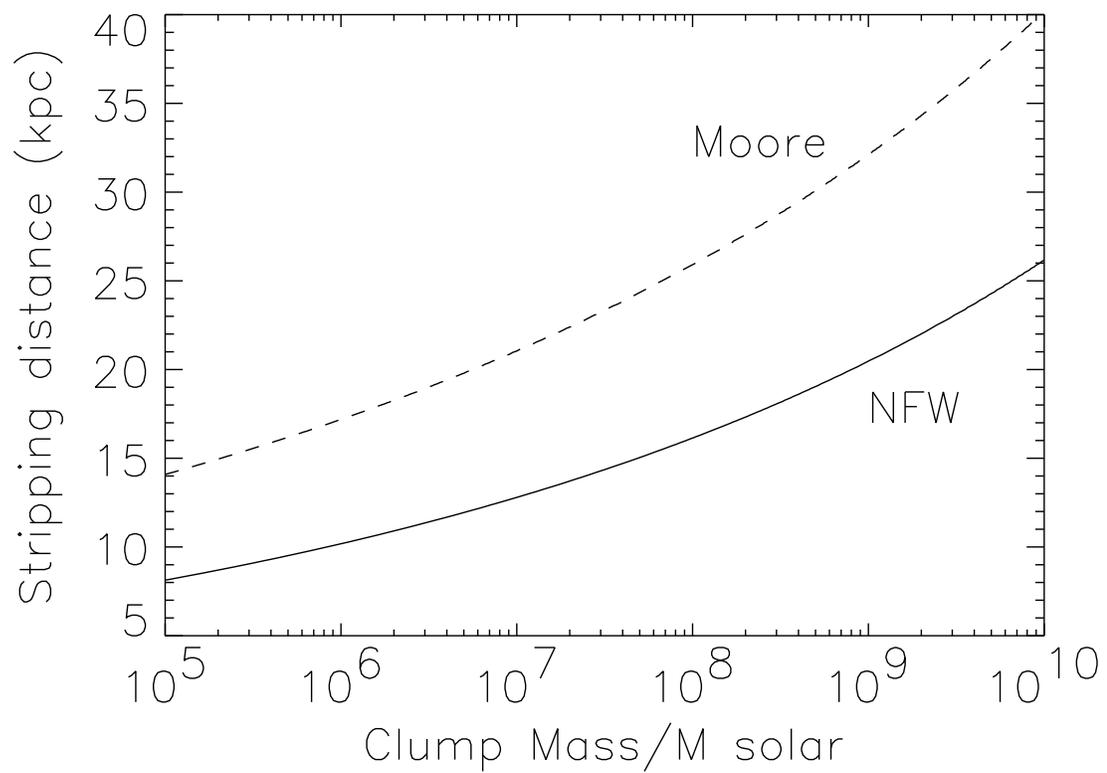}
\caption{Stripping radius of clumps as a function of the clump mass, for NFW
and Moore density profiles.}
\label{fig:strip}
\end{figure}

Using the concentration parameter plotted in fig. \ref{fig:conc} for NFW
clumps, we can estimate a stripping distance, which we define as the
distance from the galactic center where the density of a clump at the
fiducial  radius $r_f$  equals  the density in the smooth halo. At the 
stripping distance,  layers of the clump outside the fiducial
radius are stripped off by the Galactic halo tidal forces. The stripping 
distance defined in this way is plotted in fig.
\ref{fig:strip}.

\clearpage

\clearpage

From this estimate it is easy to see that most clumps
in the inner parts of the host halo are stripped of most of their
mass, so the number of clumps in the inner regions is very low.
In fact, most clumps that are able to reach the central part of the Galaxy
merge with the smooth dark matter profile. Only relatively lower mass
older clumps can reach closer to the inner regions without significant
disruption since they are in general more concentrated and have denser cores.
Numerical simulations show the disappearance of large clumps in  the centers of
galaxy size halos, but they cannot resolve the smaller mass clumps that may
eventually make their way into the core of the galaxy. 
In order to bracket the plausible range in clump distributions, we make the
very conservative  assumption in scenario II  that the inner
10 kpc of the Galaxy halo have no clumps at all.

\section{Gamma-ray emission}

In order to determine the gamma ray emission from the DM clumps, we model the
distribution of clumps in the Galaxy with a fit to numerical simulations as
in Blasi and Sheth (2000). The probability distribution function of clumps
with a given mass and at a given position  $d$  (measured from
the Galactic center) is given by:
\begin{equation}
N_c (d,M_c) = N_{c,0} \left(\frac{M_c}{M_H}\right)^{-\alpha}
\left(1+\left (\frac{d}{d_{sc}}\right)^2 \right)^{-3/2},
\label{clumpdist}
\end{equation}
where $N_{c,0}$ is a normalization constant and $d_{sc}$
is the scale radius of the clump distribution (assumed to be 10 kpc
as in Blasi and Sheth 2000). Simulations find $\alpha\simeq 1.9$  and a
halo like that of our Galaxy, with $M_H\simeq 2\times 10^{12}~M_\odot$,
contains about $500$ clumps with mass larger than $10^8~M_{\odot}$ (Ghigna et
al. 1998).

The gamma ray flux per unit solid angle and per unit energy along a fixed
line of sight in the $(\theta,\phi)$ direction can be computed as
\begin{equation}
\Phi_{\gamma}(E_{\gamma},\theta,\phi)=\frac{1}{4\pi}\int_0^{s_{max}}
ds \int_{M_{min}}^{\zeta M_H} N_c(d(s),M) {\cal N}_{\gamma} ~dM
\label{gammaflux}
\end{equation}
where $d(s)=\sqrt{s^2-2sd_{\odot}cos\theta+d_{\odot}^2}$ is the
distance of a generic point on the line of sight from the Galactic center
(with $\theta$ as the angle between the direction $s$ and the axis Sun-Galactic
center), $M$ is the clump mass, $\zeta M_H$ is the maximum allowed mass
for DM clumps in the Halo. We use $\zeta=0.01$, since the Milky Way halo 
does not show recent mergers of satellites with masses larger than $\sim 2\times 10^{10}~M_\odot$ (the Large Magellanic Cloud has about $9 \times 10^{9}~M_\odot$).  
${\cal N}_{\gamma}$
is the total number of photons emitted per unit time and energy by a DM
clump of mass $M$. This quantity, depending on the scenario chosen for the
clump density profile, may depend or not on the distance $d(s)$ of the
considered point from the Galactic center. In the first scenario, where the
normalization of the clump density is related to the smooth Halo density,
 ${\cal N}_{\gamma}={\cal N}_{\gamma}(M,d,E_{\gamma})$,
while in the second scenario
${\cal N}_{\gamma}={\cal N}_{\gamma}(M,E_{\gamma})$.
Note that eq. (\ref{gammaflux}) is an average over all
possible realizations of a halo with its clumpy structure. Fluctuations
around this value may be present due to the accidental proximity of a few
clumps in the specific realization that we happen to experience in our
Galaxy.

\clearpage

\begin{figure}[t!]
\plotone{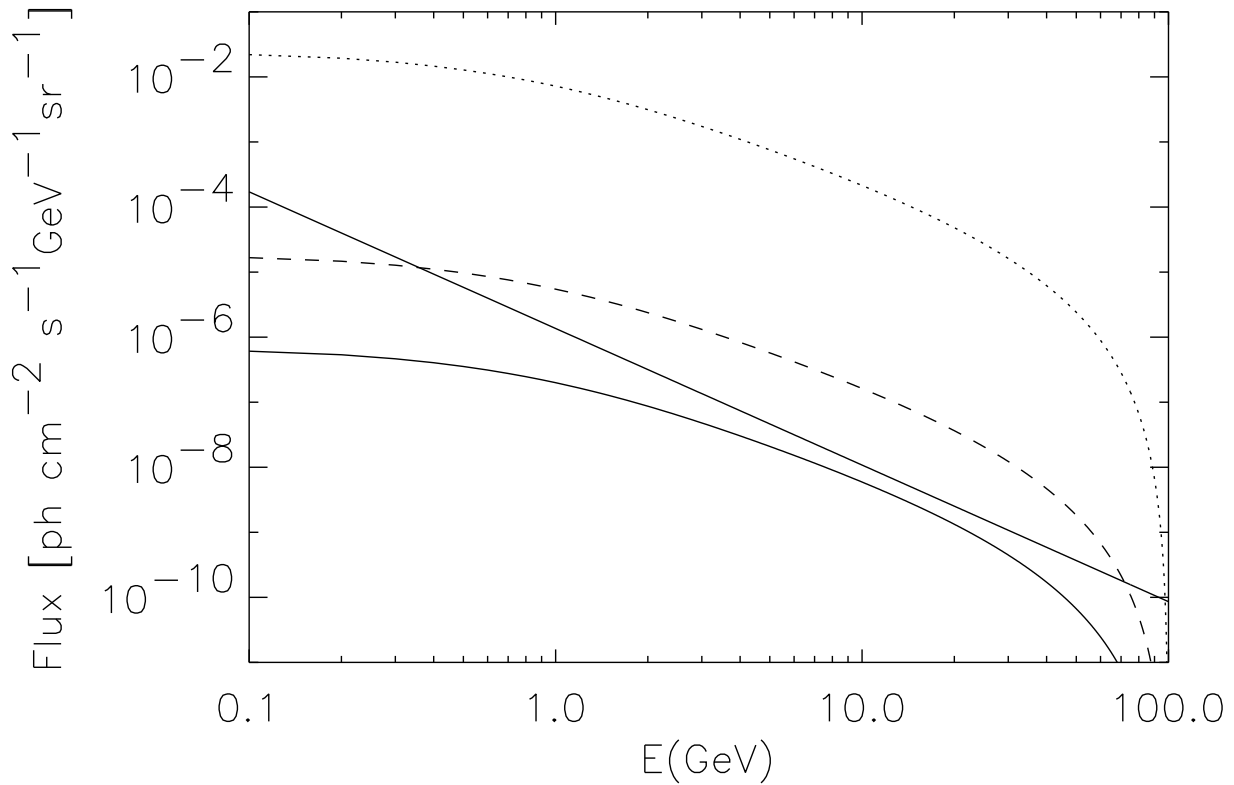}
\caption{Type I scenario. Flux of gamma-rays in units of
(GeV cm$^2$ s sr)$^{-1}$ arriving on Earth averaged in all directions for
$m_{\chi}=$ 100 GeV, $\langle \sigma v \rangle=3\times 10^{-27}$ cm$^3$/s
and $M_{c,min}=10^5$ $M_{\odot}$. SIS density profile (dotted line), Moore
et al. density profile (dashed line) and NFW density profile (continuous
line).  Also shown is the EGRET data on extragalactic diffuse gamma-ray
background
(Sreekumar 1998)}
\label{fig:fluxI}
\end{figure}

\clearpage

In principle, the neutralino annihilation production of gamma-rays has a
complex spectrum with contributions from a number of channels (for a recent
example, see Cesarini et al. 2003). Given that the uncertainties in the clump
concentration induce very large variations in the total gamma-ray flux, we
chose to calculate the gamma-ray spectrum by considering the dominant
gamma-ray emission due to $\pi^0$ production from quark-antiquark pairs and
neglected leptonic and bosonic channels. With this simplifying assumption we
derive  gamma-ray spectra that are in good agreement with Ullio et al. (2002) and
Cesarini et al. (2003) for $m_{\chi}$ less than the W-boson mass and for  
$m_{\chi}$ above the top quark mass. The discrepancy between the simplified
spectra we derive below and the fit by Berstrom et al. (2001) vary slightly
with neutralino parameters, but stay within a factor of a few for neutralino
masses from 10 GeV up to a few TeV.

The number of photons produced by quark-antiquark pairs with energy
$E_{\gamma}$ in a single
$\chi \bar{\chi}$ annihilation can be written as follows:
\begin{equation}
\frac{dN_{\gamma}}{d E_{\gamma}}=\int_{E_{\pi,min}}^{E_{\pi,max}}
dE_{\pi} P(E_{\pi},E_{\gamma})\frac{dN_{\pi}}{dE_{\pi}}
\end{equation}
where $P(E_{\pi},E_{\gamma}) = 2 (E_{\pi}^2 - m_{\pi}^2)^{-1/2}$ is
the probability per unit energy to produce a $\gamma$-ray with energy
$E_{\gamma}$ out of a pion with energy $E_{\pi}$. For the pion
fragmentation function we assume the functional form introduced in
Hill (1983):
\begin{equation}
\frac{dN_{\pi}}{dE_{\pi}} = \frac{1}{m_{\chi}}\frac{15}{16}
x^{-3/2}(1 - x)^2 \,
\label{piFF}
\end{equation}
with $x = E_{\pi}/m_{\chi}$, $E_{\pi,\rm max} = m_{\chi}$ and
$E_{\pi,\rm min} = E_{\gamma}+m_{\pi}^2/4 E_{\gamma}$. Finally,
\begin{equation}
\frac{dN_{\gamma}}{dE_{\gamma}} = \frac{5}{4 m_{\chi}} \int_{x_{\rm
m}}^{1} dx \frac{(1 - x)^2}{x^{3/2} (x^2 - \eta^2)^{1/2}}~,
\label{gamspec}
\end{equation}
where  $\eta = m_{\pi}/m_{\chi}$, and $x_{\rm m} = E_{\gamma}/m_{\chi} +
m_{\chi} \eta^2/4 E_{\gamma}$.

The neutralino annihilation rate per unit volume is,
$$\Gamma_{\chi\bar{\chi}}(r,M,E_{\gamma}) = \rho_{\chi}^2(r)
\langle \sigma v \rangle /m_{\chi}^2~,$$
therefore the $\gamma$-ray emissivity $j_{\gamma}(r,M,E_{\gamma})$
associated with the single clump of mass $M$ is obtained by multiplying eq.
(\ref{gamspec}) by $\Gamma_{\chi\bar{\chi}}$. The number of gamma rays
produced per unit time and per unit energy in a single DM clump of mass
$M$ is then
\begin{equation}
{\cal N}_{\gamma}(M,E_\gamma)=
\int_0^{r_0} dr 4\pi r^2 j_{\gamma}(r,M,E_\gamma)~.
\label{gammarate}
\end{equation}

\clearpage

\begin{figure}[t!]
\plotone{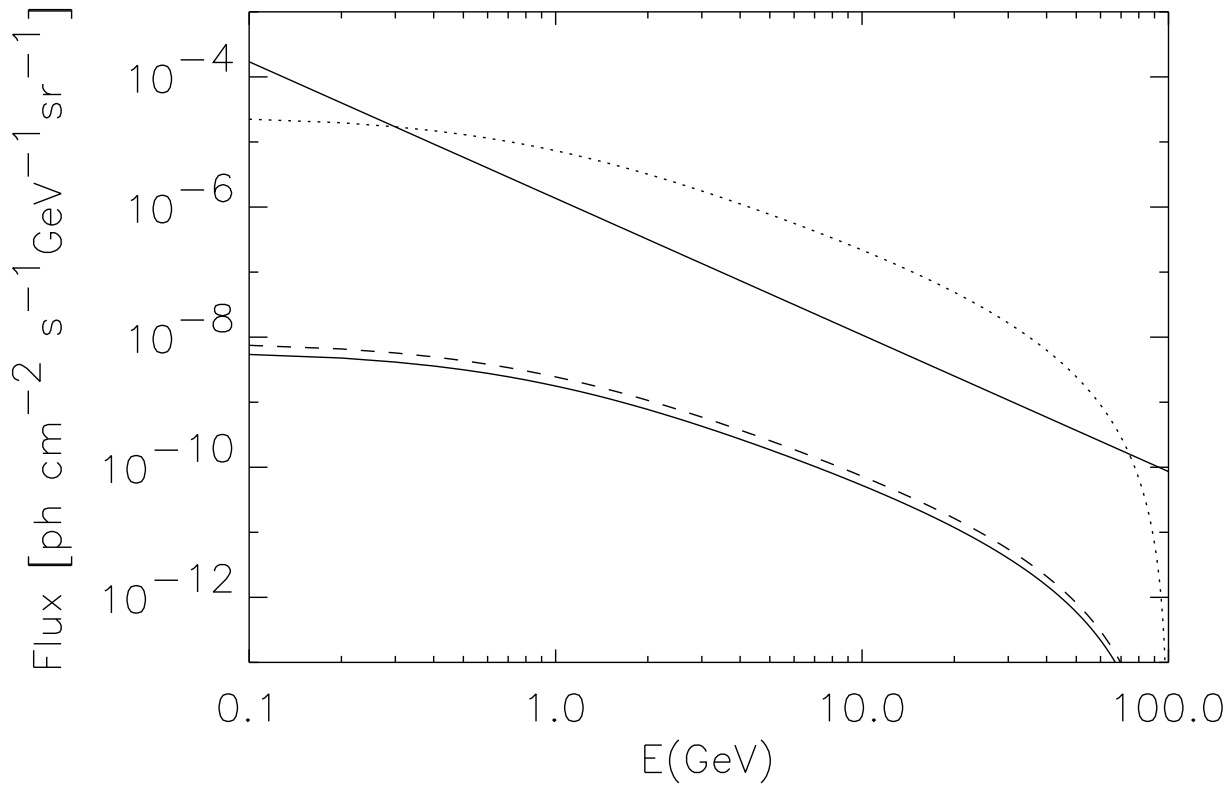}
\caption{Type II scenario. Flux of gamma-rays in units of
(GeV cm$^2$ s sr)$^{-1}$ arriving on Earth averaged in all directions for
$m_{\chi}=$ 100 GeV, $\langle \sigma v \rangle=3\times 10^{-27}$ cm$^3$/s
and $M_{c,min}=10^5$ $M_{\odot}$. SIS density profile (dotted line), Moore
et al. density profile (dashed line) and NFW density profile (continuous
line).  The solid line is the EGRET bound on extragalactic diffuse
gamma-ray background (Sreekumar 1998).}
\label{fig:fluxII}
\end{figure}

\clearpage

\section{Results}

In this section we present the $\gamma$-ray emission resulting 
from the dark matter annihilation in the halo of
our Galaxy, including both the smooth and clumped components introduced
above. We detail the description of these results for the two scenarios
(type I and II) of spatial distribution and concentration of the clumped
component. The emission in type II scenarios is much weaker than
that obtained in type I scenarios, because type
I clumps have much stronger concentration. In type II scenarios, the
gamma ray signal from clumps overcomes the gamma ray flux from the smooth
dark matter distribution only in the direction of the galactic anticenter
and only for SIS and Moore profiles for the dark matter distribution
inside the clumps. On the other hand, in type I scenarios the contribution
of the clumped  component to the diffuse gamma ray flux from the galactic
halo is several  orders of magnitude above the smooth halo component, for any
of the three  density profiles considered above. This impressive difference
in the predictions is symptomatic of a large uncertainty in the formation
and survival of dark matter substructures.

In figs. \ref{fig:fluxI} and \ref{fig:fluxII} we plot the flux of gamma-rays
in type I and II scenarios  in units of (GeV cm$^2$ s sr)$^{-1}$ arriving on
Earth averaged in all  directions for $m_{\chi}=$100 GeV and with
$\langle \sigma v \rangle = 3 \times 10^{-27}$ cm$^3$/s. The curves refer
to the gamma ray flux due to the full dark matter halo including the
smooth and the clumped components.  In both figures the dotted, dashed and
solid lines correspond to SIS, Moore and NFW clump density profiles
respectively. The fluxes are obtained for a minimum clump mass of
$M_{c,min}=10^5 M_{\odot}$. The dependence of these fluxes on the
value of $M_{c,min}$ is very weak.  The diffuse flux is the convolution of the
 mass function of the
clumps (slightly flatter than $\sim M^{-2}$)
and the gamma flux from a single clump. For the three density profiles we study, the  
 flux from a single clump scales as $\sim M$, thus the diffuse flux scales 
logarithmically with $M_{c,min}$.

For the type II scenario, the gamma-ray flux contributed by the clumped
component is comparable to the contribution of the smooth dark matter
profile, while for the type I scenario, the  clumpy component is
overwhelmingly larger than that due to the smooth component.
Also shown is the EGRET bound  on the extragalactic diffuse
gamma-ray background (Sreekumar 1998),  which is known from 30 MeV to
$\sim$ 30 GeV  to fit:
\begin{equation}
{dN_{eg} \over {d\Omega dE}}=
1.36\times 10^{-6} \left({E \over {{\rm GeV}}}\right)^{-2.10} {\rm
GeV}^{-1}{\rm cm}^{-2}{\rm s}^{-1}{\rm sr}^{-1} ~.
\label{egret}
\end{equation}

Depending on the density profile the fluxes have different scalings with
the neutralino parameters $m_{\chi}$ and $\langle \sigma v \rangle$:
\begin{equation}
\Phi_{SIS}\propto \langle \sigma v \rangle^{1/2} m_{\chi}^{-5/2}
\qquad
\Phi_{NFW,Moore}\propto \langle \sigma v \rangle m_{\chi}^{-3}.
\label{scalings}
\end{equation}
These scalings are the same for the type I and II scenarios.

\clearpage

\begin{figure}[t!]
\plotone{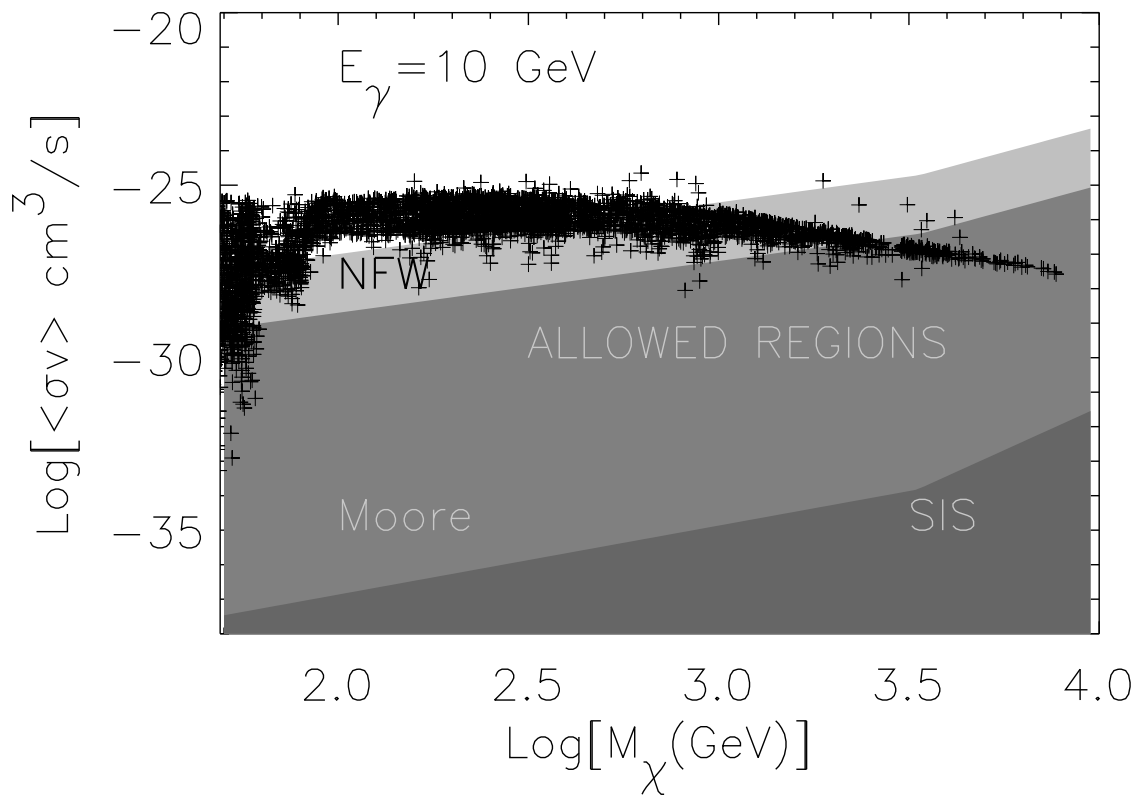}
\caption{Neutralino cross section times velocity (thermally averaged) versus neutralino mass. Crosses are neutralino models from Tasitsiomi and Olinto (2002). Shaded regions are allowed regions of neutralino parameters for type I scenarios for clumps with the labeled dark matter profiles (NFW, Moore, SIS). Models with parameters above the shaded regions are ruled out by the EGRET diffuse flux in type I scenarios.}
\label{fig:bound}
\end{figure}

\clearpage

It is clear from figs.  \ref{fig:fluxI} and \ref{fig:fluxII} that the
comparison between our predictions and the observed diffuse background
is meaningful only for the type I scenario, with highly concentrated
clumps. In the second scenario, the fluxes are too low, with the
exception of the case in which the density profile is the extreme SIS case.
For the other profiles, the region of parameters that can be constrained
is already ruled out from accelerator experiments $m_{\chi}\ge 50$ GeV
(Hagiwara et al. 2002) and from theoretical arguments
$\langle \sigma v \rangle \le 10^{-26}$cm$^{3}$/s (Baltz et al. 2002).

The situation is different for type I scenarios. As anticipated,
all the density profiles imply gamma-ray fluxes comparable to or largely
in excess of the EGRET observations of the diffuse gamma ray background.
These observations may therefore be used to extract 
constraints on the neutralino parameter space.
In fig. \ref{fig:bound}, we show the thermally averaged cross 
section times velocity of neutralinos as a function of neutralino mass.
To give a sense of where neutralino parameters tend to lie in this 
graph we plot in crosses a set of $\langle \sigma v \rangle$ versus 
$m_\chi$ generated in DarkSUSY (P.Gondolo et al 2002) 
by Tasitsiomi and Olinto (2002).
These models span a range of parameters in minimal supersymmetric 
extensions of the Standard Model (MSSM) and a range of cosmological 
dark matter densities consistent with observations. Note that the neutralino 
parameter space can generate an even smaller range of 
$\langle \sigma v \rangle$ versus 
$m_\chi$ for further constrained versions of MSSM, such as the 
constrained MSSM (CMSSM) where 
the neutralino masses is less than 500 GeV (Ellis et al. 2003) for 
a large range of MSSM parameters.

The shaded regions in fig. \ref{fig:bound} show the allowed regions of 
parameters for different clump profiles in the type I scenario for clump concentration.
Using the scalings given in eq. (\ref{scalings}) and EGRET data at 10
GeV, we find that SIS clumps in our halo are ruled  out in the region
$\langle \sigma v \rangle \ga 2 \times 10^{-35}
(m_{\chi}/{\rm 100 \, GeV})^{2} {\rm cm}^3/{\rm s}$
for 50 GeV $\leq m_\chi \leq 3 $TeV, and
$\langle \sigma v \rangle \ga 6 \times 10^{-40}
(m_{\chi}/{\rm 100 \, GeV})^{5} {\rm cm}^3/{\rm s}$
cm$^3$/s for $m_{\chi} \ga 3$ TeV.

Moore et al. clumps are also strongly constrained: the 10 GeV 
EGRET data require that
$\langle \sigma v \rangle \la 2.5 \times 10^{-28}
(m_{\chi}/{\rm 100 \, GeV})^{3/2} {\rm cm}^3/{\rm s}$ 
for 50 GeV $\leq m_\chi \leq 3$ TeV, and
$\langle \sigma v \rangle \la
10^{-30} (m_{\chi}/{\rm 100 \, GeV})^{3} {\rm cm}^3/{\rm s}$ 
for $m_{\chi} \ga 3$ TeV.
If we extend eq. (\ref{egret}) from 30 GeV to
100 GeV assuming the same functional form, 
the bounds get tighter: for $m_{\chi}$ between 50 GeV and 30
TeV, the flux from clumps is below the $\gamma$-ray background if
$\langle \sigma v \rangle \la 6 \times 10^{-29} (m_{\chi}/{\rm 100
GeV})^{3/2}$  cm$^3$/s, while for $m_{\chi} \ga 30$ TeV, the region
$\langle \sigma v\rangle \la  10^{-32} (m_{\chi}/{\rm 100 GeV})^{3}$
cm$^3$/s is allowed.

\clearpage

\begin{figure}[t!]
\plotone{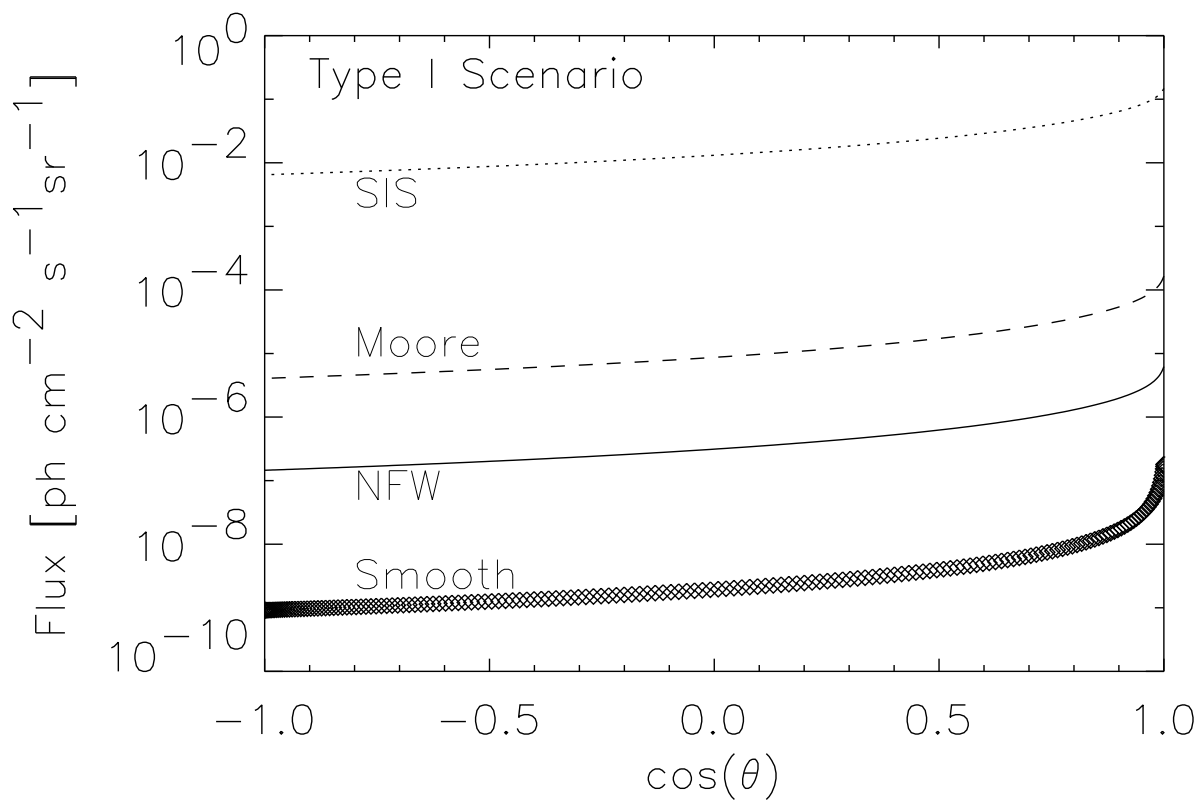}
\caption{Dependence of the energy integrated gamma ray flux on the angle
$\theta$ for the type I scenario. The solid, dashed and dotted lines refer
to clumps with NFW, Moore and SIS density profiles respectively. The thick
line is the contribution of the smooth component. A neutralino mass of 100
GeV and annihilation cross section of $3\times 10^{-27}$ cm$^3$/s have
been used.}
\label{fig:aniI}
\end{figure}

\clearpage

NFW clumps in the type I scenario are only weakly
constrained. The bounds that can be placed by EGRET at 10 GeV are as follows:
if $m_{\chi}$ is between 50 GeV and 3 TeV, the allowed region is defined by
$\langle \sigma v \rangle \la  10^{-26} (m_{\chi}/{\rm 100
GeV})^{3/2}$  cm$^3$/s. For $m_{\chi} \ga 3$ TeV, the allowed region is
instead
$\langle \sigma v \rangle \la 6 \times 10^{-29} (m_{\chi}/{\rm 100 GeV})^{3}$
cm$^3$/s. If we extrapolate eq. (\ref{egret}) from 30 GeV to 100 GeV, the
bounds become: for $m_{\chi}$ between 50 GeV and 30 TeV, one must
have $\langle \sigma v \rangle \la 3 \times 10^{-27} (m_{\chi}/{\rm 100
GeV})^{3/2}$  cm$^3$/s while for $m_{\chi} \ga 30$ TeV, the allowed region
becomes
$\langle \sigma v\rangle \la 5 \times 10^{-31} (m_{\chi}/{\rm 100 GeV})^{3}$
cm$^3$/s. All these bounds are shown in fig. \ref{fig:bound}.

\clearpage

\begin{figure}[t!]
\plotone{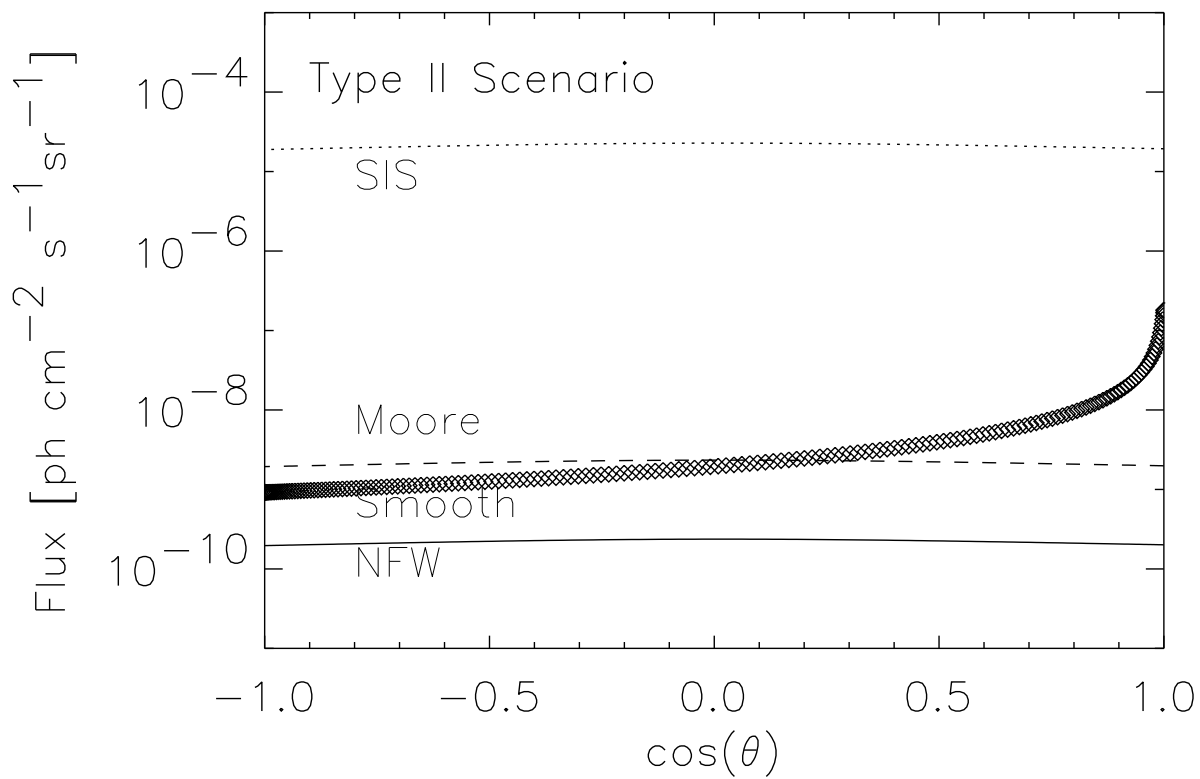}
\caption{Dependence of the energy integrated gamma ray flux on the angle
$\theta$ for the type II scenario. The solid, dashed and dotted lines refer
to clumps with NFW, Moore and SIS density profiles respectively. The thick
line is the contribution of the smooth component. A neutralino mass of 100
GeV and annihilation cross section of $3\times 10^{-27} $ cm$^3$/s have
been used.}
\label{fig:aniII}
\end{figure}

\clearpage

An important signature of the diffuse gamma ray emission from dark
matter  annihilation in the halo of our Galaxy is the anisotropy due to the
off-center position of the solar system compared to the center of the dark
matter distribution.
The anisotropy for the type I and type II scenarios is shown in
figs. \ref{fig:aniI} and \ref{fig:aniII} respectively, where we
have plotted the energy integrated flux above $100$ MeV versus
$\cos(\theta)$ (equal to 1 in the direction of the galactic center).
In these plots an isotropic flux would appear as a flat line.
The energy integrated flux on the y-axis is defined as
\begin{equation}
I(\theta)=\frac{1}{2\pi}
\int_{E_{\gamma,min}}^{m_{\chi}} dE_{\gamma}
\int_0^{2\pi} d\phi \Phi_{\gamma}(E_{\gamma},\theta,\phi),
\label{anisot}
\end{equation}
where $\Phi_{\gamma}(E_{\gamma},\theta,\phi)$ is obtained from eq.
(\ref{gammaflux}).

 From fig. \ref{fig:aniI} one can see that a large anisotropy is predicted
for the type I scenario. The curves plotted there refer to the case of
$m_{\chi}=100$ GeV and $\langle \sigma v \rangle=3\times 10^{-27}$cm$^3$/s.
The solid, dashed and dotted lines refer to NFW, Moore and SIS profiles
respectively, while the thick curve is the contribution to the gamma ray
flux due to the smooth component in the direction defined by $\cos \theta$.

In fig. \ref{fig:aniII}, we plot the same curves for the scenario of type II.
The anisotropies obtained in this case are clearly smaller than the
previous case, as indicated by the almost flat lines. This is due to our
conservative assumption that there are no clumps in  the inner 10 kpc of the
Galaxy. In this case, the contribution of the clumps to the gamma-ray flux is
never dominant over the smooth  component for clumps with NFW
profile, while for the SIS clumps emission is always dominant. For clumps with
Moore profile, the clumpy component overcomes the smooth component in the
region  around the Galactic anticenter.

\section{Conclusion}

The annihilation of dark matter in the halo of our galaxy generates a
diffuse background of gamma-rays. The dark matter substructure of the
galactic halo can be a dominant component of the diffuse gamma-ray flux
depending on the concentration and location of DM clumps in the inner regions
of the Galactic halo.

In order to bracket the range of possible fluxes, we considered two
extreme scenarios, that we named type I and type II. The first corresponds
to  extremely concentrated clumps present everywhere in the Galaxy halo,
while in the second scenario the clumps are much less concentrated and are
completely destroyed by tidal effects in the inner 10 kpc of the Galaxy. While
the type I scenario allows one to put very strong constraints on the
properties of neutralinos and on the density profile inside clumps, the type
II scenario generates  fluxes of diffuse gamma-rays which are  barely
detectable, with the exception  of the case in which the density profile of
the clumps is modeled as a SIS sphere.

For the type I scenario, most of the parameter space of neutralino dark matter
is ruled out if the density profile of dark matter clumps is in the form of a
SIS sphere or a Moore profile. Weaker bounds can be imposed on the neutralino
parameter space in the case of NFW density profile.

Recent N-body simulations have started to gather evidence of the destruction
of clumps in the inner regions of dark matter halos, which  
favor type II like scenarios. On the other hand, numerical simulations
only resolve relatively high mass clumps and the distribution of more
concentrated smaller clumps is yet to be understood, although there is some 
recent progress in this direction (Berezinsky, Dokuchaev and Eroshenko 2003).
Finally,  one should keep in mind the possibility of detecting isolated 
nearby clumps with Atmospheric Cherenkov telescopes (Tasitsiomi and
Olinto 2002), which  may be more promising than looking at diffuse fluxes.

\section*{Acknowledgments}

We thank A. Kravtsov, F. Stoehr, and R. Wechsler for many helpufull
discussions. This work was supported in part by the Center for Cosmological
Physics at the University of Chicago through grant NSF PHY-0114422, by the
NSF through grant AST-0071235 and DOE grant DE-FG0291-ER40606 at the
University of Chicago, and by the Italian Minister of Research through Cofin
2002.

\end{document}